Modeling Homophily in Dynamic Networks with Application to HIV Molecular Surveillance

Victor DeGruttola,  Department of Biostatistics, Harvard TH Chan School of Public Health, Visiting Scientist, UCSD

Masato Nakazawa, Cytel, Inc.  Waltham MA

Jinyuan Liu,  Division of Biostatistics and Bioinformatics, Herbert Wertheim School of Public Health and Human Longevity Science, UC San Diego

Xin Tu, Division of Biostatistics and Bioinformatics, Herbert Wertheim School of Public Health and Human Longevity Science, UC San Diego

Susan Little, Division of Infectious Diseases and Global Public Health, University of California San Diego.

Sanjay Mehta, Veterans Affairs, San Diego Healthcare System, San Diego, CA, USA

Corresponding author: Victor DeGruttola. vdegruttola@gmail.com.
USPS Mailing address:  PO Box 1041, Truro, MA 02666

Abstract
This paper describes a novel approach to modeling homphily, i.e. the tendency of nodes that share (or differ in) certain attributes to be linked; we consider dynamic networks in which nodes can be added over time but not removed. Our application is to HIV genetic linkage analysis that has been used to investigate HIV transmission dynamics.  In this setting, two HIV sequences from different persons with HIV (PWH) are said to be linked if the genetic distance between these sequences is less than a given threshold.  Such linkage suggests that that the nodes representing the two infected PWH, are close to each other in a transmission network; such proximity would imply that either one of the infected people directly transmitted the virus to the other or indirectly transmitted it through a small number of intermediaries.  These  viral genetic linkage networks are dynamic in the sense that, over time, a group or cluster of genetically linked viral sequences may increase in size as new people are infected by those in the cluster—either directly or through intermediaries. Our approach makes use of a logistic model to describe homophily with regard to demographic and behavioral characteristics—that is we investigate whether similarities (or differences) between PWH in these characteristics impacts the probability that their sequences are be linked.  Such analyses provide information about HIV transmission dynamics within a population.

**Introduction**

Below we describe a novel approach to modeling homphily, i.e. the tendency of nodes in a dynamic network that share features to be linked, applied to the setting of HIV molecular surveillance. In this setting, two HIV sequences from different persons with HIV (PWH) are said to be linked if the genetic distance between these sequences is less than a given threshold. Such linkage suggests that that the nodes representing the two infected PWH, are close to each other in a transmission network, in the sense that either one of the infected people directly transmitted the virus to the other or indirectly transmitted it through a small number of intermediaries[1]. Such analyses are intended to help direct field resources in ways that can best contain this pandemic. Our statistical approach makes use of a logistic model for the analysis of homophily. This flexible model allows for consideration of the extent to which similarities and differences in characteristics (demographic, behavioral, biological) impact viral genetic linkage between HIV genetic sequences and are therefore are likely to impact HIV transmission. The viral genetic linkage networks are dynamic in the sense that, over time, a group or cluster of genetically linked viral sequences may increase in size as new people are infected by those in the cluster; as mentioned above, the transmission can be direct or through intermediaries[2].

Such considerations regarding homophily in networks are directly related to one of the four "pillars" (Diagnose, Protect, Treat, and Respond) of the plan for Ending the HIV epidemic in the US as outlined by the Centers for Disease Control (CDC); the Respond pillar focuses on monitoring and then acting on outbreaks by providing prevention and treatment services[3]. Molecular epidemiology, which relies on genetic linkage described above, is now a key tool to monitor ongoing HIV epidemics for clusters of related transmissions, and to direct responses to these potential outbreaks[4-6]. A limitation of this approach is that linked infections between people can only be observed among PWH or their contacts if the latter can be identified. However, public health responses associated with viral genetic linkage analyses are often coupled with HIV partner notification to provide prevention services to persons who may be living with HIV infection and unaware or vulnerable to HIV. Therefore, identifying characteristics that cause newly-infected people to link to existing genetic clusters could help in identifying persons who are at high risk of acquiring HIV in the future—and thereby in guiding provision of biomedical prevention resources, such as pre-exposure prophylaxis (PrEP) to them. Amirkhanian noted that "Network interventions are feasible and powerful for reducing unprotected sex and potentially for increasing HIV testing uptake" [7].

Modeling homophily to investigate features of transmission dynamics requires identifying the characteristics (e.g., age, race/ethnicity, neighborhood of residence) that tend to link a newly infected individual with specific viral genetic clusters of PWH — either because such characteristics are similar in this collection of people (an example of homophily) or because they are different (an example of heterophily). Both can occur within the population under study. For example, some newly linked young people might tend to link preferentially to clusters of people of their age, whereas others might link preferentially to clusters of older people. This could lead to a bimodal distribution in the difference between the age of newly linked individuals and the mean age of the people in the cluster—and hence a bimodal distribution in the ages of people in clusters themselves. We discuss how to model such possibilities below.

An example of identification of homophily in an HIV transmission network arose in the demonstration of greater viral genetic linkage among Black PWH who have the same income levels compared to those with different income levels[8]. In addition, vulnerability to HIV infection among Black men who have sex with men has been observed to increase when individuals enter high-risk sexual networks characterized by high density and racial homogeneity;[7,9] such behavioral dynamics might be expected to result in homophily in transmission networks. Similarly among persons who inject drugs (PWIDs), homophily by ethnicity[10] and injecting behaviors[11] has been observed. Homophily may also provide insights regarding transmission of other infections. For example, SARS-CoV-2 has spread more rapidly in certain neighborhoods and certain ethnic/racial and social groups[12]—which may have resulted from homophily in transmission networks[13]. In Japan, Andalibi et al. showed that viral transmission networks of SARS-CoV-2 demonstrated age homophily, as well as homophily between symptomatic and asymptomatic cases, possibly suggesting a virologic effect on transmission[14]. Groups of people who share characteristics (e.g., membership in a church group[16], vacationed at a ski resort[17]) also may be more likely to transmit to each other–either in single transmission or superspreader events. In another example, sexually transmitted infections (STIs) were shown to have been transmitted at greater rates between partners of similar education status in an analysis of five African cities[15]. These examples highlight how knowledge of homophily and heterophily, such as would be revealed in analyses using our methods, could provide insights about transmission dynamics.

As is the case for COVID-19, clusters of HIV infection tend to grow at highly variable rates[18]. Associations between characteristics of newly linked individuals and HIV viral genetic clusters may strengthen, stabilize, or weaken over time; our logistic model can accommodate such phenomena by treating the relevant parameters as time varying, though we do not investigate this possibility in our illustrative example. Below we describe the use of a logistic model to describe homophily of demographic and behavioral characteristics in a dynamic HIV transmission network to provide information about their impact on probability of forward transmission.

**STUDY POPULATION**

Between July 1, 1996 and March 31, 2018, ART-naïve adult and adolescent (≥16 year-olds) PWH were prospectively recruited to observational research studies at the University of California San Diego (UCSD). For details, see Little et al. (CID, 2021)[18]. Data collected at the baseline visit included: HIV genotype (partial *pol* sequence), testing for bacterial sexually transmitted infections (STIs) (gonorrhea [GC], chlamydia [CT], and syphilis), and routine labs needed for clinical care. Baseline participant characteristics are presented in **Table 1.**

**HIV NETWORK INFERENCE**

Population HIV partial *pol* nucleotide consensus sequences were derived for all participants (GenoSure® MG, LabCorp Specialty Testing Group, South San Francisco or Viroseq v.2.0; Celera Diagnostics, Alameda, CA). If more than one HIV sequence was available for a participant, only the earliest was included in this analysis. We inferred the HIV network by computing all pairwise genetic distances between partial *pol* sequences from each participant (i.e., network node) and connected nodes for which the corresponding genetic distance was less than 1.5% using HIV-TRACE (hereafter, the UCSD method). For further details and information about accessing sequences, see Little et al.[18,19] To create the HIV network, Little et al.[18] identified PWH--i.e., nodes--that did not link to any earlier nodes in the network, which began in 1996. These were defined as "seeds" and followed over time. For each seed or cluster that arose from a seed, we counted the number of incident nodes that subsequently linked to that seed or cluster.

**NETWORK HOMOPHILY ANALYSIS**

In following each seed--or cluster that grew from a seed--we define an event of linkage to be the event that the $i^{th}$ PWH (henceforth $PWH_i$) joins any of the clusters available when the linkage takes place. We define the time of the event of linkage $t_i$ to be the date of first sequence of $PWH_i$. We model the probability of this event using the following custom logistic equation

$$\pi_{ij_i} = \frac{exp(\sum_{k=0}^{K} x_{(ij_i)k} \beta_k)}{1 + exp(\sum_{k=0}^{K} x_{(ij_i)k} \beta_k)}$$

where *i* indexes the individual PWH; $j_i$ denotes a cluster that was formed before the sequence date of $PWH_i$ ($t_i$); *k* indexes the predictor to be included in the model; and *x* and *β* denote predictors and parameters, respectively. We refer to each $PWH_i$ as a newly linked case (NLC) at time $t_i$.

We let $y_{ij_i} = \{0,1\}$, where 1 indicates that the $PWH_i$ joined cluster $j_i$ and 0, that $PWH_i$ did not join that cluster. The probability mass function for a Bernoulli distribution is

$$P(Y = y_{ij_i} | X = x_{ij_i}) = \pi_{ij_i}^{y_{ij_i}} \cdot (1 - \pi_{ij_i})^{(1-y_{ij_i})}$$

To estimate parameters $\beta_k$ and hence $\pi_{ij_i}$, we computed likelihood of all data using the following likelihood function:

$$\begin{aligned} L(\beta) &= \prod_{j_i}^{J_i} P(Y = y_{ij_i} | X = x_{ij_i}) \\ &= \prod_{j_i}^{J_i} \pi_{ij_i}^{y_{ij_i}} \cdot (1 - \pi_{ij_i})^{(1-y_{ij_i})} \end{aligned}$$

where $J_i$ is the number of clusters that are available for $PWH_i$ to join at time $t_i$ (i.e. clusters that were formed before $t_i$).

To test the hypothesis that homophily is associated new cluster linkage, we create a homophily variable for different types of characteristics of PWH. For binary outcomes, like Hispanic Ethnicity (HE). We model:

$$x_{(ij_i)} = r_{j_i}^{x_i}(1 - r_{j_i})^{x_i-1}$$

where $r_{j_i}$ is the proportion of cluster members with HE for cluster $j_i$, and $x_{(ij_i)} = 1$ if $PHW_i$ is positive for HE and 0, if negative. When computing this proportion, the the NLC is not included in the cluster membership. We also define a homophily variable for the absolute value of the difference in age between the NLC and the average age of the members of clusters $j_i$ for $j_i = 1, 2, \ldots, J_i$. This variable is calculated as the absolute value of the difference in birth year between $PWH_i$ ($BY_i$) and the mean for each of the clusters $j_i$ ($\overline{BY_{j_i}}$), $j_i = 1, \ldots, J_i$ at time $t_i$. Thus, our homophily variable, denoted birth year difference (BYD), is defined as $BYD = |BY_i - \overline{BY_{j_i}}|$.

Figure 1 provides a histogram of the differences in age between the newly linked cases and the clusters to which they were linked. The plot shows that the newly linked cases tended to be younger than those in the cluster of linkage; the 25% and 75% percentiles of this distribution are -6.0 and 1.75. We also note that the plot is unimodal. Figure 2 displays the boxplots for this difference for clusters of different maximum size. No strong relationship between cluster size and this distribution is evident in this figure. In order to accommodate the possibility that linkage could increase with both small and large values of BYD compared to values in the middle range, we can also use functions of it—for example quadratic—in the model above.

For diseases gonorrhea, chlamydia, and syphilis, STI categorical homophily variables were created based on **Table 2**, which defines three categories of homophily: positive, neutral and negative. In this table, *r* is the proportion of cluster members that are STI positive; once again, the NLC was excluded from calculation of *r*. For this analysis, cases with the neutral homophily category were excluded. The reason for this choice is that homophily is harder to interpret in settings whether clusters are mixed in STI status.

Hypothesis tests of the null hypothesis that the homophily variable has the null value (does not impact risk of joining particular clusters) is based on a likelihood ratio test. We first consider univariate models to examine whether each predictor was associated and then include those with p-value<0.05 in a multivariable model including STI individually and then jointly. Confidence intervals are obtained from the Fisher information (see supplementary note).

**RESULTS**

Baseline participant characteristics are presented in Table 1. Age and ethnicity were available for all participants; but, as indicated in the table, there was a fairly large group of individuals for whom STI information was not available.

HOMOPHILY AND SOCIODEMOGRAPHIC CHARACTERISTICS

As shown in Table 3, there was strong positive homophily associated with hispanic ethnicity (HE), and strong negative homophily, with birth year difference (BYD). The second result implies that the larger the difference between the age of the NLC and the average age for an available cluster, the lower the odds of the NLC joining that cluster. In addition, there was a significant interaction between BYD and HE on the odds of linkage. The results from the multivariable model imply that with BYD=0 and when NLC links to a single PWH, the odds of linkage increases by a factor of 3.90 (95% CI 2.86, 5.37) if the NLC and PWH available for linkage share the same HE compared to when they differ. If the NLC links to a cluster of two people of different HE, the odds of linkage is $\sqrt{3.90}=1.97$ compared to when neither share HE with the NLC. The table also shows that for two people of the same HE, the linkage odds ratio decreases by a factor of 0.90 (0.88, 0.93) for each additional year of difference in BYD. There was a significant interaction between BYD and HE for negative homophiliy; the odds ratio (95% confidence interval) associated with BYD by HE interaction was 0.93 (0.89, 0.96), $p<0.001$. This result imples that that for NLC and PWH with the same HE, the odds ratio associated with BYD effect is 0.90 x 0.93= 0.84—which is close to the univariate effect. We also included a quadratic as well as linear effect of BYD along with HE in a model; the quadratic effect was nearly 0 and was associated with a high p-value.

The Hosmer Lemershow test for the multivariable model implies a reasonable fit (Chi-square statistic = 8.57, df = 4, p-value = 0.0729). However, the first two percentile groups—obtained, as is traditional for this test, from the ordered values of estimated probabilities of NLC joining a single PWH or a cluster--had relatively few observed linked events (5 and 37, respectively, Supplemental Table 1). Collapsing these two categories, yielded a Hosmer Lemershow test that showed stronger support for the model fit (Chi-square statistic = 5.83, df = 3, p-value = 0.1204, Supplemental Table table 2, BYD and HE were strongly associated with negative homophiliy and positive homophily respectively. The odds ratio (95% confidence interval) associated with BYD was 0.86 (0.84, 0.87), $p<0.001$ for both univariate and multivariable models.

HOMOPHILY AND SEXUALLY TRANSMITTED INFECTIONS.
Table 4 shows the frequency distribution of linkages by homophily type (postive, neutral, or negative) and specific sexualy transmitted disease. The upper panel of Table 5 shows that, when investigated individually, none of the STI homophily variables impacted the probability of linkage. While the current syphillis indicator had a relatively high odds ratio (1.55), the small number of study participants in this category (28) provided limited power; and the 95% confidence interval did not exclude the null value. An additional homophily variable was considered: the presence or absence of any STI; once again no significant effect was observed.  These effects remained qualitatively the same after adjustment for HE and difference in BYD, although the estimated odds ratio for syphillis is somewhat reduced.

**Discussion**

Our proposed method allows for longitudinal evaluation of homophily in dynamics networks; our particular focus is is on newly identified cases of HIV infection that genetically link to clusters of HIV infected individuals.  The method could apply to any other dynamic network in which ties are created or dissolved over time.  To incorporate dissolution of ties we could consider a polytimous logistic regression model in which events of both linkage and dissolution of linkage are modeled. Our approach depends on construction of homophily variables; as we demonstrate, these can be quite general. Here we analyzed variables of different types to illustrate the flexibility of the approach. Knowledge of how characteristics of newly linked cases of HIV infection impact probability of joining clusters with particular characteristis provides useful information about transmission dynamics.  The homophily variables may consider both similarity and dissimilarity in these characteristics—and both types of variables should be considered.  We know that both homophily and heterophily may be present and could be detected though choices of model or through evaluation of fit.  One example is the situation in which  some people preferentially select partners based on similarity of age, and others based on difference in age. Knowledge of the sociological factors relevant for constructing homophily variables may be useful—as is in-depth investigation of the types of patterns observed in the data.  For example, preferences for similarity or difference in age may reflect other demographic characteristics.

Understanding of transmission dynamics can aid in targetting prevention resources. For example, knowing the features of individuals that may make them more likely to join certain clusters, because they share (or are dissimilar) in those features, could help prioritize prevention resources to people in clusters with characteristics that make them most likely to sustain future growth from linked incident infections.   These characteristics may be defined by clinical, demographic, and other factors. Similarly, knowing the features of those most likely to join growing clusters may also help in prioritizing PrEP.  Together, the knowledge of the clinical and demographic factors associated with growing clusters and the factors associated with persons linking to those clusters provides a blueprint for how to direct limited prevention resources in the most efficient manner.

**Table 1. Baseline Participant Characteristics**

| Number of Participants | N=1119 |
|---|---|
| Race/Ethnicity; n (%) | |
|     White (non-Hispanic) | 560 (50.0) |
|     Black (non-Hispanic) | 89 (8.0) |
|     Hispanic | 341 (30.5) |
|     Other/Unknown | 129 (11.5) |
| Birth Year; median (IQR) | 1973 (1965,1982) |
| Gonorrhea; n (%)[1] | 49 (6.6) |
| Chlamydia; n (%)[1] | 62 (8.4) |
| Syphilis; n (%)[1] | 28 (3.8) |
| Clustered; n (%) | 532 (47.5) |

1: STI: Gonorrhea, Chlamydia and Syphilis were not assessed for 377, 378, and 374 participants.

**Table 2**. Definition of Homophily Status for STIs

| Infection Status | Proportion r of Cluster members that are STI positive | Homophily Status |
|---|---|---|
| Non-Infected | 0<r<100 | Neutral |
| Non-Infected | r=0 | Positive |
| Non-Infected | r=100 | Negative |
| Infected | 0<r<100 | Neutral |
| Infected | r=0 | Negative |
| Infected | r=100 | Positive |

r is the proportion of cluster members (excluding the newly linked case) that are STI positive

**Table 3 Logistic Model Results with All Cases**

| Effect | Univariable Models | | | Multivariable Models | | |
|---|---|---|---|---|---|---|
| | OR | OR 95% CI | p | OR | OR 95% CI | p |
| Abs(ΔBY) | 0.86 | (0.84, 0.87) | <0.001 | 0.90 | (0.88, 0.93) | <0.001 |
| Hispanic | 2.44 | (1.99, 3.02) | <0.001 | 3.90 | (2.86, 5.37) | <0.001 |
| Abs(ΔBY)* Hispanic | | | | 0.93 | (0.89, 0.96) | <0.001 |

**Table 4. Frequency distribution of linkages by homophily type and sexualy transmitted disease**

| Infection | Cumulative Cluster Proportion | Categorical Homophily | Chlamydia | Gonorrhea | Syphilis | Any STI |
|---|---|---|---|---|---|---|
| Negative | 0<r<100 | Neutral | 212 | 218 | 91 | 163 |
| Negative | r=0 | Positive | 404 | 424 | 588 | 391 |
| Negative | r=100 | Negative | 6 | 3 | 1 | 5 |
| Positive | 0<r<100 | Neutral | 51 | 34 | 18 | 100 |
| Positive | r=0 | Negative | 36 | 31 | 12 | 49 |
| Positive | r=100 | Positive | 1 | 0 | 0 | 2 |

**Table 5. Logistic Model Results without Neutral Cases**

| Effect | Univariable Models | | |
|---|---|---|---|
| | OR | OR 95% CI | p |
| Gonorrhea | 0.90 | (0.66, 1.26) | 0.53 |
| Chlamydia | 0.88 | (0.63, 1.28) | 0.49 |
| Active Syphilis | 1.55 | (0.93, 2.83) | 0.12 |
| Any STI | 1.11 | (0.84, 1.49) | 0.49 |

| | Multivariable Models | | | | | |
|---|---|---|---|---|---|---|
| | Chlamydia | | | Gonorrhea | | |
| Effect | OR | OR 95% CI | p | OR | OR 95% CI | p |
| STI | 0.9 | (0.66, 1.27) | 0.54 | 0.87 | (0.62, 1.25) | 0.42 |
| Abs(ΔBY) | 0.81 | (0.79, 0.83) | <0.001 | 0.81 | (0.79, 0.83) | <0.001 |
| Hispanic | 3.29 | (2.52, 4.37) | <0.001 | 3.47 | (2.65, 4.60) | <0.001 |
| | Syphilis | | | Any STI | | |
| Effect | OR | OR 95% CI | p | OR | OR 95% CI | p |
| STI | 1.27 | (0.76, 2.33) | 0.4 | 1.07 | (0.81, 1.44) | 0.66 |
| Abs(ΔBY) | 0.83 | (0.82, 0.85) | <0.001 | 0.81 | (0.79, 0.83) | <0.001 |
| Hispanic | 2.77 | (2.20, 3.52) | <0.001 | 3.28 | (2.51, 4.35) | <0.001 |

Abbreviations: OR: odds ratio, p: p-value; Abs(ΔBY): the absolute value of the birth-year difference, CI: confidence interval, STI: sexually transmitted infection.

**Supplemental Table 1. Hosmer Lemershow Test for the Multivariable Model with 6 Bins.**

| P(Y=1) | Observed (Y=0) | Observed (Y=1) | Predicted (Y=0) | Predicted (Y=1) |
|---|---|---|---|---|
| [7.37e-08, 0.000832] | 23144 | 5 | 23139 | 10.02 |
| (0.000832, 0.00196] | 24286 | 37 | 24289 | 33.82 |
| (0.00196, 0.00334] | 22082 | 73 | 22096 | 58.71 |
| (0.00334, 0.00548] | 22589 | 107 | 22598 | 97.74 |
| (0.00548, 0.0097] | 24155 | 177 | 24154 | 177.56 |
| (0.0097, 0.0234] | 21509 | 308 | 21488 | 329.14 |

**Supplemental Table 2. Hosmer Lemershow Test for the Multivariable Model with 5 Bins (the first two sparse bins was collapsed).**

| P(Y=1) | Observed (Y=0) | Observed (Y=1) | Predicted (Y=0) | Predicted (Y=1) |
|---|---|---|---|---|
| [7.37e-08, 0.00196] | 47430 | 42 | 47428 | 43.85 |
| (0.00196, 0.00334] | 22082 | 73 | 22096 | 58.71 |
| (0.00334, 0.00548] | 22589 | 107 | 22598 | 97.74 |
| (0.00548, 0.0097] | 24155 | 177 | 24154 | 177.56 |
| (0.0097, 0.0234] | 21509 | 308 | 21488 | 329.14 |

Supplementary Note

To get the information matrix, we used the score (formula 14) and the Hessian (formula 15).

$$\mathbf{u}(\underline{\beta}) = \mathbf{X}^T(\underline{y} - \underline{\pi}) \quad (14)$$

$$\frac{\partial^2 LL(\underline{\beta})}{\partial \beta_k \partial \beta_l} = \underline{x_k}\underline{y} - \underline{x_k}\underline{x_l}\underline{\pi}(1-\underline{\pi}) \quad (15)$$

-

**References**

1. Wertheim JO, Kosakovsky Pond SL, Forgione LA, et al. Social and Genetic Networks of HIV-1 Transmission in New York City. *PLoS Pathog.* 2017;13(1):e1006000.
2. Novitsky V, Steingrimsson J, Howison M, et al. Longitudinal typing of molecular HIV clusters in a statewide epidemic. *AIDS.* 2021.
3. CDC. Respond. *Ending the HIV Epidemic* https://www.cdc.gov/endhiv/respond.html. Accessed April 4, 2021, 2021.
4. Wertheim JO, Panneer N, France AM, Saduvala N, Oster AM. Incident infection in high-priority HIV molecular transmission clusters in the United States. *AIDS.* 2020;34(8):1187-1193.
5. Oster AM, France AM, Panneer N, et al. Identifying Clusters of Recent and Rapid HIV Transmission Through Analysis of Molecular Surveillance Data. *J Acquir Immune Defic Syndr.* 2018;79(5):543-550.
6. Tumpney M, John B, Panneer N, et al. Human Immunodeficiency Virus (HIV) Outbreak Investigation Among Persons Who Inject Drugs in Massachusetts Enhanced by HIV Sequence Data. *J Infect Dis.* 2020;222(Suppl 5):S259-S267.
7. Amirkhanian Y, Kelly J, Kuznetsova A, et al. Using social network methods to reach out-of-care or ART-nonadherent HIV+ injection drug users in Russia: addressing a gap in the treatment cascade. *J Int AIDS Soc.* 2014;17(4 Suppl 3):19594.
8. Kobe J, Talbot O, Chen I, et al. Short Communication: Viral Genetic Linkage Analysis Among Black Men Who Have Sex With Men (HIV Prevention Trials Network 061). *AIDS Res Hum Retroviruses.* 2019;35(5):434-436.
9. Janulis P, Phillips G, Birkett M, Mustanski B. Sexual Networks of Racially Diverse Young MSM Differ in Racial Homophily But Not Concurrency. *J Acquir Immune Defic Syndr.* 2018;77(5):459-466.
10. Aitken CK, Higgs P, Bowden S. Differences in the social networks of ethnic Vietnamese and non-Vietnamese injecting drug users and their implications for blood-borne virus transmission. *Epidemiol Infect.* 2008;136(3):410-416.
11. Wu Z, Detels R, Zhang J, et al. Risk factors for intravenous drug use and sharing equipment among young male drug users in Longchuan County, south-west China. *AIDS.* 1996;10(9):1017-1024.
12. Cyrus E, Clarke R, Hadley D, et al. The Impact of COVID-19 on African American Communities in the United States. *Health Equity.* 2020;4(1):476-483.
13. Furuse Y, Sando E, Tsuchiya N, et al. Clusters of Coronavirus Disease in Communities, Japan, January-April 2020. *Emerg Infect Dis.* 2020;26(9).
14. Andalibi A, Koizumi N, Li MH, Siddique AB. Symptom and Age Homophilies in SARS-CoV-2 Transmission Networks during the Early Phase of the Pandemic in Japan. *Biology (Basel).* 2021;10(6).
15. Kenyon C, Colebunders R. Birds of a feather: homophily and sexual network structure in sub-Saharan Africa. *Int J STD AIDS.* 2013;24(3):211-215.
16. Voeten H, Sikkema RS, Damen M, et al. Unravelling the modes of transmission of SARS-CoV-2 during a nursing home outbreak: looking beyond the church super-spread event. *Clin Infect Dis.* 2020.
17. Danis K, Epaulard O, Benet T, et al. Cluster of Coronavirus Disease 2019 (COVID-19) in the French Alps, February 2020. *Clin Infect Dis.* 2020;71(15):825-832.
18. Little SJ, Chen T, Wang R, et al. Effective HIV Molecular Surveillance Requires Identification of Incident Cases of Infection. *Clin Infect Dis.* 2021.
19. Little, S. J., Kosakovsky Pond, S. L., Anderson, C. M., Young, J. A., Wertheim, J. O., Mehta, S. R., . . . Smith, D. M. (2014). Using HIV networks to inform real time prevention interventions. PLoS ONE, 9(6), e98443. doi:10.1371/journal.pone.0098443

**Figure 1**

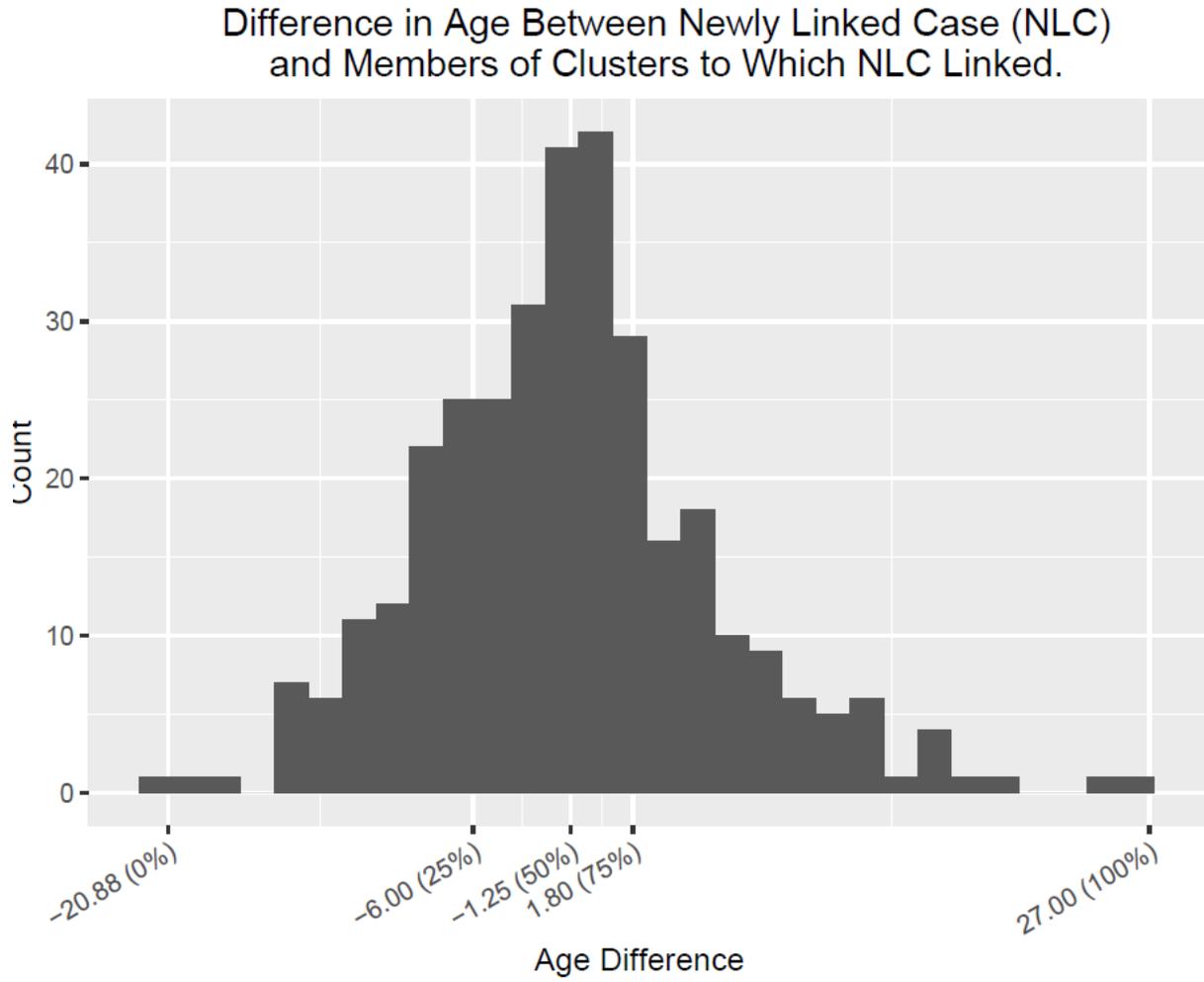

Figure 2

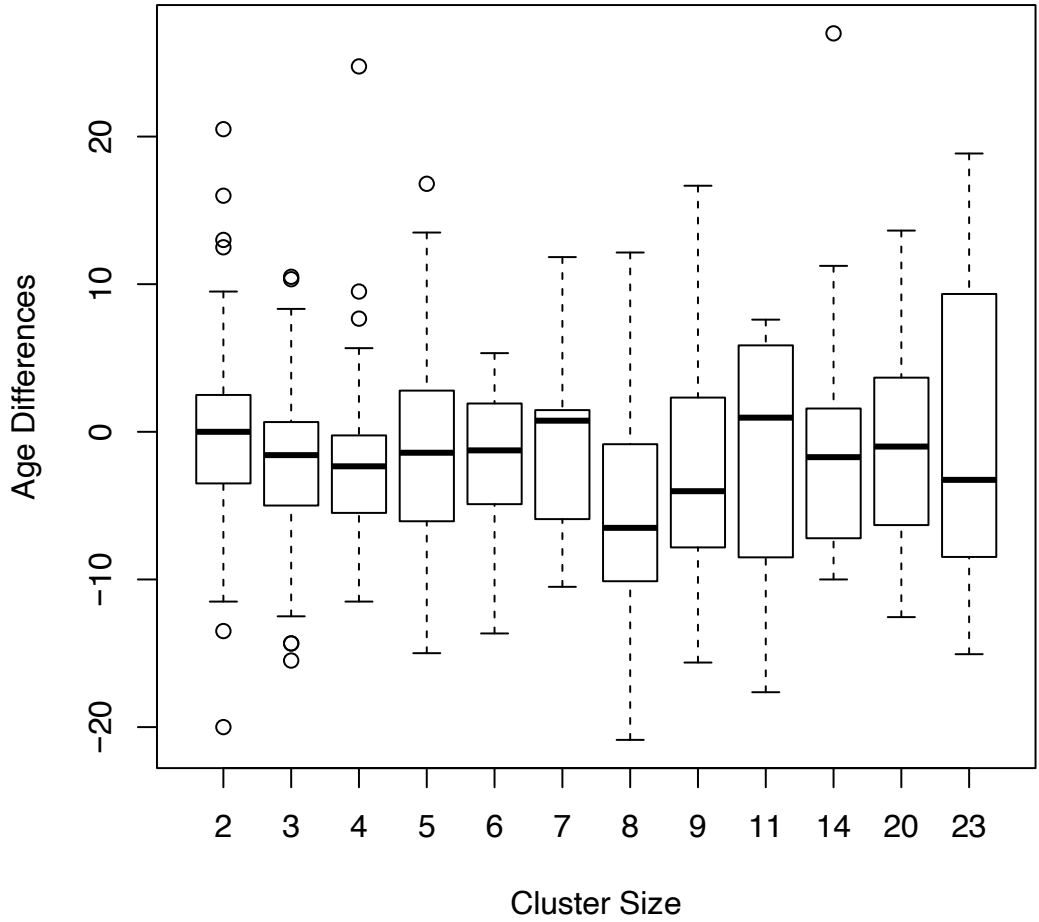

**Acknowledgements:** NIH R24 AI106039-05, NIH R37 AI51164 (DeGruttola), NIH CFAR AI036214.